\newcommand{\TODO}[1]{}
\newcommand{\bvec}[1]{\mathbf{#1}}
\newcommand{\order}[1]{\mathcal{O} \left( #1 \right)}
\documentclass[
 aip,
 jmp,
 amsmath,amssymb,
reprint,%
letterpaper,
]{revtex4-1}

\usepackage{graphicx}
\usepackage{dcolumn}
\usepackage{bm}
\usepackage{hyperref}
\usepackage{amsmath}
\usepackage{amssymb}

\begin{document}

\title[Determining U(1) charges in WCFFHS models]{Algorithm for determining U(1) charges in\\ free fermionic heterotic string models}

\author{William Hicks}
\email[]{William\_Hicks@Brown.edu}
\altaffiliation{Department of Physics, Brown University, Providence, RI 02912, USA}
\affiliation{CASPER, Department of Physics, Baylor University, Waco, TX, 76798-7316, USA}

\author{Lesley Vestal}
\email[]{lvestal@ubc.interchange.ca} 
\altaffiliation{Department of Physics and Astronomy, University of British Columbia, Okanagan, Kelowna, British Columbia V1V 1V7, Canada}
\affiliation{CASPER, Department of Physics, Baylor University, Waco, TX, 76798-7316, USA}

\author{Jared Greenwald}
%\email[]{} 
\affiliation{CASPER, Department of Physics, Baylor University, Waco, TX, 76798-7316, USA}

\author{Douglas Moore}
%\email[]{}
\affiliation{CASPER, Department of Physics, Baylor University, Waco, TX, 76798-7316, USA}

\author{Timothy Renner}
%\email[]{}
\affiliation{CASPER, Department of Physics, Baylor University, Waco, TX, 76798-7316, USA}

\author{Gerald Cleaver}
%\email[]{}
\affiliation{CASPER, Department of Physics, Baylor University, Waco, TX, 76798-7316, USA} 

\date{\today}

\begin{abstract}
To assist in the search for phenomenologically realistic models in the string landscape, we must develop tools for investigating all gauge charges, including $U(1)$ charges, in string models. We introduce the process for constructing fermionic string models and present an algorithm for determining the $U(1)$ gauge states and $U(1)$ charges in weakly-coupled free fermionic heterotic string (WCFFHS) models given their matter and non-Abelian gauge content. We determine the computational complexity of this algorithm and discuss a particular implementation that can be used in conjunction with a framework developed at Baylor University for building WCFFHS models. We also present preliminary results regarding matter state uniqueness for a run of 1.4 million gauge models and find that $U(1)$ charges affect the number of unique matter states in a significant minority of models constructed. We conclude by considering future avenues of investigation to which this algorithm may be applied.
\end{abstract}


\maketitle

\section{Introduction}

String theory models have the potential to completely describe both the particle content and field interactions of our universe, yet attempts to obtain precise, testable predictions based on string theory have been stymied by the sheer number of viable models. Work by a number of theorists in the early 2000's helped establish that the number of stable models, though finite, could be upwards of $10^{100}$  [\cite{bousso2000, ashok2004, douglas2003}]. Any predictions based on string theory will require a more complete understanding of this collection of models, known as the string theory landscape. In particular, systematic searches of the landscape can help us gain a statistical understanding of the range of available models. Such searches, however, require fast algorithms for determining properties of string models. In this paper, we discuss a new algorithm used to determine the $U(1)$ charges of particles in weakly coupled free fermionic heterotic string (WCFFHS) models. Such models are of particular interest due to a number of studies which have discovered phenomenologically realistic models of this type \cite{cleaver1999, lopez1992, faraggi1989, faraggi1992, antoniadis1990, leontaris1999, faraggi1991, faraggi1992_2, faraggi1992_3, faraggi1991_2, faraggi1991_3, faraggi1995, faraggi1996, cleaver1997, cleaver1997_2, cleaver1997_3, cleaver1998, cleaver1998_2, cleaver1998_3, cleaver1998_4, cleaver1998_5, cleaver1999_2, cleaver1999_3,  cleaver2000, cleaver2000_2, cleaver2001, cleaver2001_2, cleaver2002, cleaver2002_2, cleaver2002_3, perkins2003, perkins2005, cleaver2008, greenwald2009, cleaver2011}.

\subsection{WCFFHS model building\label{sec:intro:model_building}}

For our investigation of $U(1)$ charges, we built on research conducted at Baylor University to develop a fast, versatile framework for building WCFFHS models \cite{renner2011}. The process of model-building begins with a set of 64-component basis vectors which specify information about the boundary conditions for fermions on the world-sheet as well as a valid GSO (Gliozzi, Scherk, and Olive) projection matrix, which is used to eliminate non-physical particle states encountered during model-building. The 64 components of the basis vectors are divided into two parts, the first of which specifies information about the left-movers on the world sheet and the second of which specifies information about the right-movers. The number of components in each part is determined by the number of large spacetime dimensions. For four large spacetime dimensions, the left-moving part consists of 20 components while the right-moving part consists of 44 components. The 64 components of the entire vector are usually labeled as shown:
\begin{align}
&\left( \psi^{{1,2}} (x,y, w)^{{1,...,6}} || \right. \nonumber \\
&\left. \bar{\psi}^{1,...,10} \bar{\eta}^{1,...,6} \bar{y}^{1,...,6} \bar{w}^{1,...,6} \bar{\phi}^{1,...,16}\right).
\end{align}
Additionally, each input vector has an order $N$, which specifies the denominator of all components in the basis vector. These components take on values between $-1$ and 1.

These vectors must obey a number of constraints to constitute valid basis vectors. One particularly useful constraint specifies that every component in the entire vector must be paired with an equal component in order for the vector to be valid. This allows us to represent these vectors in a complex basis using only half as many components, an important advantage for manipulating them in calculations. In the description of our algorithm, we will use a complex basis to describe all vectors.

The input basis vectors must also obey certain constraints to guarantee that the generated model satisfies the condition of modular invariance. This condition ensures that modular transformations on the world-sheet will not alter the physics of the model. The specific constraints on the input basis vectors $\{\vec{\alpha}_i^B\}$ with orders $\{N_i\}$ are given by
\begin{eqnarray}
N_i \left(\vec{\alpha}_i^B\right)^2 &=& 0 \pmod{16} \textrm{ for even N} \\
N_i \left(\vec{\alpha}_i^B\right)^2 &= &0 \pmod{8} \textrm{ for odd N}\\
N_{ij} \vec{\alpha}_i^B  \cdot \vec{\alpha}_j^B &= &0 \pmod{8},
\end{eqnarray}
where $N_{ij}$ is the least common multiple of $N_i$ and $N_j$.

Given a set of valid basis vectors, we construct sectors $\{\vec{\alpha}_i\}$ by taking all linear combinations of the form
\begin{equation}\label{eqn:sectors}
\vec{\alpha} = \sum m_j \alpha_j^{B}
\end{equation}
where the $m_j$ obey
\begin{equation} \label{eqn:m_j}
m_j  \in \mathbb{Z} \cap [0,N_j).
\end{equation}
The sectors determine how the free world-sheet fermions $f_i$ transform when transported around a non-contractible loop as given by
\begin{equation}
f \longrightarrow e^{-i\pi\alpha} f.
\end{equation} 
This information is determined by the topology of the world-sheet. All possible charges or particle states $Q^{\vec{\alpha}}$ may then be constructed from these sectors as given by
\begin{equation}
Q^{\vec{\alpha}}= \frac{1}{2}\vec{\alpha} + \vec{F},
\end{equation}
where the components of the fermion number operator $\vec{F}$ take on values given by
\begin{equation}
\vec{F}_p \in \{1, 0, -1\}.
\end{equation}

We maintain only states that are massless at the string scale, because these are the states that exist at low energies and therefore represent observable particles. We exclude string-scale massive particles because they will not be observable. Additionally, we apply the GSO projection to remove non-physical states such as tachyons. This projection ensures the remaining gauge states fit together to form a group theoretic root structure, compatible with the remaining matter states \cite{robinson2008}. The GSO coefficient matrix $k_{ij}$ is subject to modular invariance constraints of its own, given by \cite{renner2011}
\begin{eqnarray}
N_jk_{ij} &=  &0  \pmod{2} \\
k_{ij}+k_{ji} &= & \frac{1}{2}\vec{\alpha}_i^B \cdot \vec{\alpha}_j^B \pmod{2} \\
k_{ii}+k_{i1} &= &\frac{1}{4}\vec{\alpha}_i^B - s_i  \pmod{2}.
\end{eqnarray}
Given a valid GSO coefficient matrix, we reject any particle states failing the requirement
\begin{equation}\label{eqn:6}
\vec{\alpha}_i^{B} \cdot \vec{Q}^{\alpha} =\sum m_j k_{ij} + s_i \pmod{2}
\end{equation}
where $s_i$ equals 0 if $\alpha_i^{B}$ is a bosonic state, and 1 if it is fermionic, and $m_j$ and $\alpha$ are defined as in \autoref{eqn:sectors}.

We can identify the type of particle that a particular state specifies by considering the components of the left-moving part for that state. Specifically, for fermions
\begin{equation}
\psi^1 = \psi^2 = \pm\frac{1}{2},
\end{equation}
and for gauge bosons
\begin{equation}
\psi^1 = \psi^2 = \pm1,
\end{equation}
with sign indicating chirality. Furthermore, all other left-moving components of gauge states are equal to zero. Thus, if the right-moving part of the $U(1)$ gauge states can be generated, adding the left-moving part is straightforward. Once all non-physical states have been removed from the model, we can determine the $U(1)$ charges of the remaining physical matter states.

\subsection{$U(1)$ charges in WCFFHS models}

In order to fully understand the gauge interactions in a particular string model, we must determine the $U(1)$ charges of the matter states. This point is perhaps best illustrated by considering the role of $U(1)$ charges in the familiar context of the Standard Model. In the Standard Model, the charge of a particle under $U(1)$ determines that particle's hypercharge, which in turn determines its electric charge through the Gell-Mann-Nishijima formula \cite{gottfried1984}
\begin{eqnarray}
Q = I_3 + \frac{Y}{2},
\end{eqnarray}
where $Q$ is the electric charge of the particle, $I_3$ is an eigenvalue of its isospin, and $Y$ is its hypercharge.

In addition to completing the picture of the gauge interactions for a model, analyzing $U(1)$ charges can help eliminate unwanted exotic particles from a model - in particular fractionally charged exotic particles. If a certain type of $U(1)$ charge known as an anomalous $U(1)$ charge is present in a model, we can break that $U(1)$, providing a mass to the exotic particles at the string scale and thereby eliminating them from the low-energy effective field theory \cite{cleaver1999_2, cleaver1998_5}. This can help dramatically in searching for phenomenologically realistic models, but identifying models of this type requires an efficient method for calculating $U(1)$ charges.

\section{Algorithm for determining $U(1)$ charges \label{sec:algorithm}}

In order to determine the $U(1)$ charges for the matter states in a string model, we must first construct the $U(1)$ gauge states. These gauge states, by definition, have a number of convenient mathematical properties that simplify their construction. Firstly, because they are gauge states, their left moving parts are already determined, as noted in \autoref{sec:intro:model_building}. Secondly, we know that the right-moving part of the $U(1)$ states must be orthogonal to the right-moving parts of the simple roots of the non-Abelian gauge groups in the model. If the state were not orthogonal to the simple roots of any one of the non-Abelian gauge groups, it would be a part of that gauge group and not the $U(1)$ gauge group. Finally, since representations of $U(1)$ have only a single generator, every $U(1)$ gauge group in the model has exactly one gauge state and hence, the right-moving parts of every $U(1)$ gauge state must be orthogonal to the right-moving part of every other $U(1)$ gauge state. Thus, if we can construct a set of vectors which are orthogonal to the right-moving parts of the simple roots of the non-Abelian gauge groups in a model as well as to each other, we will have constructed the right-moving parts of all $U(1)$ gauge states in the model. Once the gauge states have been constructed, we take the dot product of the right moving part of each of the matter states with the right-moving part of the $U(1)$ gauge states to determine its charge under each $U(1)$ in the model.

Throughout our discussion of the algorithm for generating $U(1)$ gauge states, we will use a complex basis for all vectors, as described in \autoref{sec:intro:model_building}. So, we can represent every pair of equal components in the real basis, using a single component in the complex basis. To simplify our discussion of the problem, we denote the set of vectors given by the right-moving parts of the simple roots of the non-Abelian gauge groups by $\mathbf{V}_{SR}$. We denote the vectors whose components are given by the right-moving parts of the $U(1)$ gauge states by $\mathbf{V}_{U(1)}$. Furthermore, we denote the cardinality of $\mathbf{V}_{SR}$ by $N_{SR}$ and the cardinality of $\mathbf{V}_{U(1)}$ by $N_{U(1)}$\TODO{Figure out why this linebreak isn't happening.}.\\

\subsection{Outline of the algorithm}

Our algorithm for generating $U(1)$ gauge states begins with the set $\mathbf{V}_{SR}$ for a particular model. The dimension $D_{RM}$ of vectors in this set is given by
\begin{eqnarray}
D_{RM} &=& 26 - D ,
\end{eqnarray}
where $D$ denotes the number of large spacetime dimensions for the model \cite{renner2011}. Since the vectors of $\mathbf{V}_{SR}$ are formed from the simple roots of gauge groups, they are by definition linearly independent. Furthermore, since $\mathbf{V}_{U(1)}$ contains all vectors orthogonal to the vectors of $\mathbf{V}_{SR}$ and to each other, $\mathbf{V}_{SR} \cup \mathbf{V}_{U(1)}$  forms a complete basis. Therefore,
\begin{eqnarray}
N_{U(1)} &=& D_{RM} - N_{SR}.
\end{eqnarray}
If $D_{RM} = N_{SR}$, then the model in question contains no $U(1)$ gauge groups.

We start by attempting to generate any vectors in $\mathbf{V}_{U(1)}$ whose components all have value 0 except for one component which takes on the value 1. This can easily be done by constructing a matrix whose rows correspond to the components of the vectors in $\mathbf{V}_{SR}$ and performing Gauss-Jordan elimination on that matrix. After elimination, we identify all columns containing only 0's, and we generate vectors in $\mathbf{V}_{U(1)}$ which consist of all 0's except for the component corresponding to one of these columns. Gauge states of this form are known as external gauge states\TODO{Is this the correct terminology?}. We then proceed to generate any remaining vectors of $\mathbf{V}_{U(1)}$.

Let $n_{U(1)}$ denote the number of vectors in $\mathbf{V}_{U(1)}$ which have already been generated, where $n_{U(1)}$ may range from $0$ to $N_{U(1)}$. In order to construct an additional vector $\bvec{v}$ of $\mathbf{V}_{U(1)}$ from $\mathbf{V}_{SR}$ and the $n_{U(1)}$ previously generated vectors, we must ensure that the dot product of $\bvec{v}$ with every vector in $\mathbf{V}_{SR}$ as well as with every previously generated vector in $\mathbf{V}_{U(1)}$ is equal to $0$. Since there are $D_{RM}$ components, this gives us a system of $N_{SR} +n_{U(1)}$ linear constraint equations on $D_{RM}$ unknowns. This means that the system has $f$ free variables, where $f$ is given by
\begin{eqnarray}
f &=& D_{RM} - N_{SR} - n_{U(1)}.
\end{eqnarray}
Thus we, can assign arbitrary values to $f$ components of $\bvec{v}$ and then straightforwardly solve for the remaining components of $\bvec{v}$ by solving this system of equations. If we let $F$ denote the value of $f$ when $n_{U(1)}$ equals $0$, this further implies that the vectors of $\mathbf{V}_{U(1)}$ will be unique only up to scaling and rotations in $F$ dimensions.

Although we must assign arbitrary values to $f$ components of $\bvec{v}$, we are not entirely free in choosing the components to which we will be assigning values. This is perhaps most easily illustrated by considering the problem of finding a vector $\bvec{w}$ orthogonal to the following (3-dimensional) vectors:
\begin{eqnarray}
\left(\frac{1}{3},\frac{2}{3},\frac{2}{3}\right)\nonumber\\
\left(\frac{2}{3},\frac{1}{3},\frac{1}{3}\right)
\end{eqnarray}
Since there are two vectors given, the system of constraint equations will consist of two equations in three unknowns. Thus, we may arbitrarily assign a value to one component of $\bvec{w}$. Let $w_i$ denote the $i$th component of $\bvec{w}$, where $i$ ranges from 1 to 3. If we choose to assign a value of 1 to $w_1$, then (taking the dot product of $\bvec{w}$ with the given vectors), the constraint equations are
\begin{eqnarray}
\frac{2}{3} w_2 + \frac{2}{3} w_3 &=& -\frac{1}{3} \nonumber \\
\frac{w_2}{3} + \frac{w_3}{3} &=& -\frac{2}{3}
\end{eqnarray}
Attempting to solve for $w_2$ and $w_3$, we see that the constraint equations reduce to:
\begin{eqnarray}
0 &=& 3\nonumber\\
0 &=& -\frac{3}{2}
\end{eqnarray}
The system is clearly inconsistent. A quick inspection of the original vectors shows that this occurred because they were not linearly independent in the subspace formed by the components for which we were solving.

In order to avoid this problem in generating a vector $\bvec{v}$ in $\mathbf{V}_{U(1)}$, we construct a matrix whose rows are determined by the components of the vectors in $\mathbf{V}_{SR}$ as well as the vectors in $\mathbf{V}_{U(1)}$ which have already been generated. Then, we perform Gaussian elimination on that matrix. After Gaussian elimination, we identify those columns which do not contain pivots and assign values to the corresponding components of $\bvec{v}$. We can safely do this since the vectors are now guaranteed to form linearly independent subspaces in the remaining components.

To illustrate this method, consider the previous example of generating the 3-dimensional vector $w$. This time, we begin by creating a matrix from the given vectors:
\begin{equation}
\left( \begin{array}{ccc}
\frac{1}{3}&\frac{2}{3}&\frac{2}{3}\\
\frac{2}{3}&\frac{1}{3}&\frac{1}{3}
\end{array} \right)
\end{equation}
After performing Gaussian elimination, this gives
\begin{equation}
\left( \begin{array}{ccc}
2&1&1\\
0&\frac{3}{2}&\frac{3}{2}.
\end{array} \right)
\end{equation}
Since the third column of this matrix does not contain a pivot, we can safely assign a value (arbitrarily chosen to be 1) to $w_3$. Proceeding as before, the system of equations for the remaining components is given by
\begin{eqnarray}
w_1 + 2 w_2 &=& -2 \nonumber \\
2 w_1 + w_2 &=& -1 .
\end{eqnarray}
This system is consistent (as expected), and we get:
\begin{equation}
\bvec{w} \equiv (0, -1, 1)
\end{equation}

After assigning all necessary values to the components of $\bvec{v}$ and solving for the remaining components, we add $\bvec{v}$ to the set of generated vectors and repeat until we have produced the $N_{U(1)}$ vectors in $\mathbf{V}_{U(1)}$. These vectors specify the components of the right-moving parts of the $U(1)$ gauge states, and we can trivially prepend the necessary left-moving part to make the full state a gauge state. Then, taking the dot product of the right-moving part of each $U(1)$ gauge state with the right-moving part of each matter state in the model, we produce the charges of each matter state under each $U(1)$ gauge group in the model.\\

\subsection{Implementation details and additional considerations}

Our implementation of the algorithm was written as an extension of a framework developed at Baylor University to build WCFFHS models \cite{renner2011}. Given a set of valid input vectors as well as a valid GSO coefficient matrix, this framework generates the matter and non-Abelian gauge content for the model. Our algorithm  then generates the $U(1)$ charges for the generated matter states. The entire framework, including our implementation of the algorithm, is written in C++, and care is taken throughout to avoid floating numeric types in order to eliminate rounding errors. The framework was written with extensibility and readability in mind, since we hope that other researchers will eventually be able to use this software to further explore the string landscape. In order to guide future development, a few choices in the details of our implementation deserve particular consideration.

In our implementation, Gauss-Jordan elimination is used to solve for the unassigned components in the generated vectors. Since Gaussian or Gauss-Jordan elimination is required for two other steps in the algorithm, this allows for significant code re-use, but a few downsides to Gauss-Jordan elimination must also be considered. Firstly, Gaussian elimination is not a numerically stable algorithm, so we must consider the possibility that the program would overflow the C++ long integer type used to store the coefficients of the systems of equations. In order to minimize this possibility, we implemented partial pivoting, which substantially improves the stability of Gaussian elimination \cite{press1988}. We then tested the program by asking it to produce a complete orthogonal basis from a single randomly generated 22-dimensional vector. This goes beyond the worst case that our implementation would attempt to solve, since all non-Abelian gauge groups that could appear in WCFFHS models have at least two generators. After running several hundred thousand trials, we determined that so long as the absolute value of the components in the randomly generated vector remained below about 250 (a very reasonable assumption), the program never overflowed the long integer type. Furthermore, since this test went well beyond the worst case that our implementation will reasonably encounter, the absolute value of the components of the input vectors can likely range substantially higher without difficulty.

The second downside to Gauss-Jordan elimination is that it is typically somewhat slower than other common methods for solving systems of linear equations \cite{press1988}. This downside is balanced by a number of factors, including the fact that Gauss-Jordan elimination with pivoting is usually slightly more stable than most other general solution methods, and the fact that Gaussian or Gauss-Jordan elimination is required in other steps of the algorithm. Furthermore, since major goals of our implementation included readability and extensibility, the use of a single easily-understood algorithm was deemed preferable to several faster, more esoteric algorithms. Nevertheless, if future advances improve the speed of generating the matter and non-Abelian gauge content of string models to the point that the process of generating $U(1)$ charges becomes a bottleneck, other solution methods should be considered.

\section{Analysis of Computational Complexity}
\TODO{\section{Complexity and Runtime Analysis} \subsection{Analysis of Computational Complexity}}
In order to facilitate analysis of the computational complexity of the algorithm for generating all $U(1)$ gauge states, we consider each significant step in the process. Using the notation of \autoref{sec:algorithm}, this process begins with Gauss-Jordan elimination on an $N_{SR} \times D_{RM}$ matrix to find any external $U(1)$ gauge states. Gauss-Jordan elimination is generally $\order{n^2m}$ on the number of rows ($n$) and columns ($m$) in a matrix, but our implementation includes a sorting step after each elimination step in order to maintain optimal pivoting. At worst, this sort will be an $\order{D_{RM}{N_{SR}}^2}$ operation, but is likely to be closer to or better than the average case, which is an $\order{D_{RM}N_{SR}\log{N_{SR}}}$ operation. Thus, the entire operation of Gauss-Jordan elimination with pivoting will have complexity
\begin{equation}\label{eqn:ext_worst}
\order{D_{RM}{N_{SR}}^3}
\end{equation}
for the worst case.

Next, we must generate all internal $U(1)$ gauge states. If we let $N_{\text{ext}}$ denote the number of external $U(1)$ gauge states, then there will be $D_{RM} - N_{SR} - N_{\text{ext}}$ such states. For each state, we must perform Gaussian elimination to determine to which components we may assign arbitrary values. We must also perform Gauss-Jordan elimination to determine the values of the remaining components. In the first of these steps, the sorting process is simplified by the fact that the sorted results from the previous iteration can be stored and used to analyze the next vector to be constructed. Thus, the worst case for the Gaussian elimination step will be:
\begin{equation}
\order{D_{RM}\left(N_{SR}+n_{U(1)}\right)^2},
\end{equation}
where $n_{U(1)}$ is the number of previously generated $U(1)$ gauge states, including the external $U(1)$ gauge states. The Gauss-Jordan elimination step will have no such advantage. Thus, using the same analysis as before, the worst case complexity for this step will be
\begin{equation}
\order{\left(D_{RM} - \left(N_{SR} +n_{U(1)}\right)\right) \left(N_{SR} +n_{U(1)}\right)^3}.
\end{equation}
Both the Gaussian and Gauss-Jordan elimination steps will be repeated until $n_{U(1)} = D_{RM} - N_{SR}$. The complexity of this step will be at a maximum when 
\begin{equation}
N_{SR} + n_{U(1)}= \frac{3}{4}D_{RM}.
\end{equation}
 Thus the overall worst case complexity of the algorithm for generating $U(1)$ gauge states will be
\begin{equation}
\order{{D_{RM}}^4}.
\end{equation}
Therefore, the overall worst case complexity depends solely on the number of right-moving components in the states of the model, which is itself determined by the number of large spacetime dimensions.

\section{Initial Results}
In order to develop an initial understanding of the impact of $U(1)$ charges on analysis of WCFFHS models, we constructed and analyzed about 1.4 million models and their $U(1)$ charges. This was done using the framework developed at Baylor for building WCFFHS models and our implementation of the algorithm for generating $U(1)$ charges. For this run, we used only non-supersymmetric layer-1 gauge models with an order of 22 or less. Furthermore, we restricted our run to models which contain at least one copy of the Standard Model gauge group, as established by previous runs on model-building software developed at Baylor. While this is a fairly specific subset of the WCFFHS landscape, this run can help provide an initial sense of how $U(1)$ charges affect our understanding of previous searches of the landscape. We present a few of the more interesting results of this run here.

\begin{figure}
\includegraphics[width=63mm]{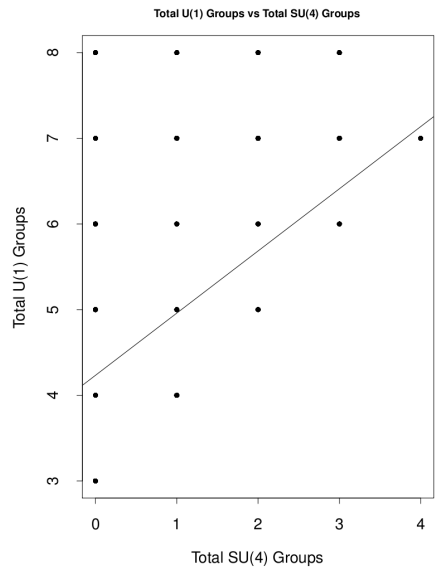}
\caption{\label{fig:U1vsSU4} A linear regression of the number of $U(1)$ groups vs the number of $SU(4)$ groups for models in our initial run}
\end{figure}

The models studied had anywhere from three to eight $U(1)$ gauge groups, with a median value of four. Since the WCFFHS model-building framework produced only the non-Abelian content of its models prior to our research, we were interested if the presence or absence of any particular non-Abelian groups was correlated with a higher number of $U(1)$ groups. Simple regression analysis showed no significant linear correlation between the number of any particular non-Abelian group in the model and the number of $U(1)$ groups in the model. In fact the highest $R^2$ value for a linear regression between the number of occurrences of a particular group and the number of $U(1)$ groups in the model was only 0.49. This result was for $SU(4)$ groups versus $U(1)$ groups as shown in \autoref{fig:U1vsSU4}.
As this figure indicates, even the correlation between $SU(4)$ and $U(1)$ groups is almost entirely insignificant.

\begin{figure}
\includegraphics[width=63mm]{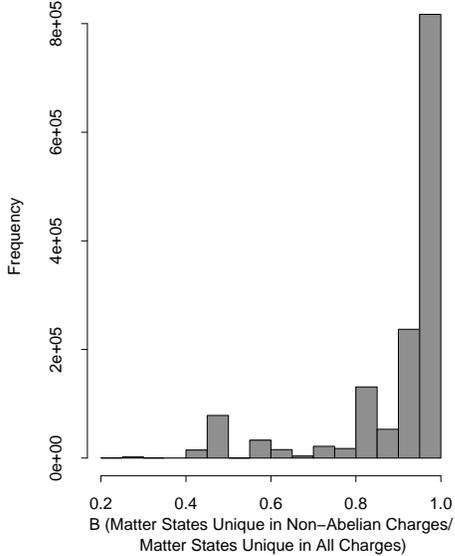}
\caption{\label{fig:B} Distribution of B for all models built}
\end{figure}

Next, we examined the role of $U(1)$ charges in determining the uniqueness of matter states within a model. While examining uniqueness between models would be of interest, the fact that $U(1)$ gauge states are unique only up to rotation prevents direct comparison of $U(1)$ charges, and related problems affect the non-Abelian charges\cite{renner2011}. Thus, we confine our analysis to comparisons of matter states within a model. In particular, we were interested in the ratio of matter states unique in their non-Abelian charges to matter states which were truly unique (i.e. unique in all charges, including $U(1)$ charges). We refer to this ratio as $B$, and it gives us some sense of how great of an error would be made in studies which only examined non-Abelian charges when attempting to identify unique matter states. The distribution of $B$ for our run of 1.4 million models is shown in \autoref{fig:B}.
As this figure indicates, consideration of $U(1)$ charges does not affect the number of unique particles for the majority of models in our study, since $B$ had a value of one for most of these models. A sizable minority had values less than one, however. Therefore, we conducted linear regression analyses of $B$ vs the occurrences of each group in our models in order to see if there were any correlations that could be used to predict the impact of $U(1)$ charges on matter state uniqueness prior to building the charges. These analyses showed no significant linear correlation between the value of $B$ and the occurrences of any given group [including $U(1)$ groups] in the model. The highest $R^2$ value among these regressions was only 0.07.

\begin{figure}
\includegraphics[width=63mm]{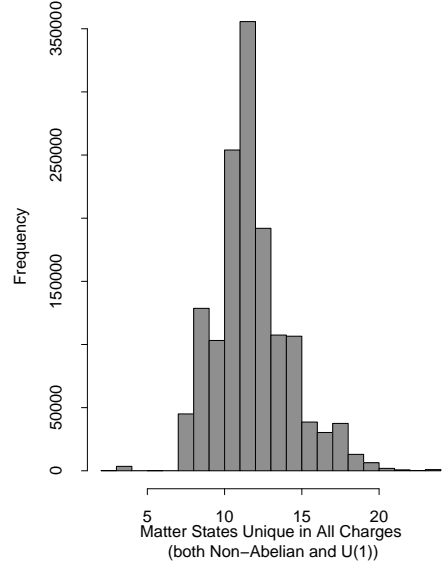}
\caption{\label{fig:UniqueParticles} Distribution of unique matter states for all models built}
\end{figure}

After examining the effect of $U(1)$ charges on matter state uniqueness, we considered the overall distribution of the number of unique matter states for the models in the run, as shown in \autoref{fig:UniqueParticles}.
As the figure indicates, this distribution is roughly symmetrical with a sharp peak at 12 unique matter states. The mean number of unique matter states was 12.27, with a standard deviation of 2.43. A Shapiro-Wilks normality test on these data gives sufficient evidence to reject the hypothesis that they follow a Gaussian distribution ($p<0.01$). It is possible, however that this lack of normality is a feature only of the specific population chosen for this study and does not hold more generally for the WCFFHS landscape. While all of these results are preliminary, they give us some sense of how the construction of $U(1)$ charges can affect our understanding of matter state uniqueness for models in the landscape.

\section{Conclusions}
In this paper, we presented a new algorithm for calculating $U(1)$ charges of particles in WCFFHS models given the matter and non-Abelian gauge content for those models. We analyzed the computational complexity, and found it to be $\order{{D_{RM}}^4}$ in the worst case, where $D_{RM}$ refers to the number of right-moving components of the particle states in a complex basis. We also discussed an implementation of this algorithm developed as an extension of a model-building framework developed at Baylor University which emphasizes readability, extensibility, and speed \cite{renner2011}, and we presented results on matter state uniqueness from an initial run of 1.4 million gauge models using this implementation. Specifically, we found that the $U(1)$ content affected the number of unique matter states only in a minority of models. We found no strong correlations that would help predict which models lie in this minority, however.

Determining the $U(1)$ charges for a particular model is important in understanding its gauge content, but it can also help in constructing phenomenologically realistic models by identifying anomalous $U(1)$ charges. Additionally, constructing $U(1)$ charges may shed some light on the problem of identifying unique models, since multiple inputs may all lead to models with the same matter and non-Abelian gauge content. While some studies have been conducted to analyze patterns in the occurrence of unique models, it is possible that by studying the $U(1)$  gauge content, we will find that some models with the same matter and non-Abelian gauge content are actually distinct due to their varying $U(1)$ charges \cite{renner2011}.

Like many aspects of the string landscape, we cannot predict everything that the $U(1)$ gauge content of string models will tell us. Only through systematic exploration of the landscape can we completely understand how $U(1)$ charges affect the search for phenomenologically realistic models.

\begin{acknowledgments}
We would like to thank the CASPER group at Baylor University for their sponsorship of the REU program under which this research was conducted. We would also like to thank the NSF for funding this research through grant PHY-1002637.
\end{acknowledgments}

\bibliography{U1_Charges_arxiv}

\end{document}